\documentclass[12pt]{article}

\usepackage[utf8]{inputenc}
\usepackage[T1]{fontenc}
\usepackage{graphicx} 
\usepackage{caption}
\usepackage{booktabs}
\usepackage{calc}
\usepackage{longtable}
\usepackage{placeins}
\usepackage{amsmath}
\usepackage{natbib}
\usepackage{hyperref}

\newenvironment{keywords}{\noindent\textbf{Keywords:}~}{\par\medskip}

\DeclareUnicodeCharacter{03B1}{\ensuremath{\alpha}}
\DeclareUnicodeCharacter{03B2}{\ensuremath{\beta}}
\DeclareUnicodeCharacter{2009}{\,}

\begin{document}

\title{The Desert Fireball Network Clear-Sky Survey}


\author{\parbox{\textwidth}{\centering
  Konstantinos S. Servis$^{1}$,
  Hadrien A. R. Devillepoix$^{1,2}$,
  Eleanor K. Sansom$^{2}$,
  Thomas W. C. Stevenson$^{1}$ \\[0.5em]
  \small $^{1}$Space Science and Technology Centre, Curtin University, GPO Box U1987, Perth WA 6845, Australia \\
  \small $^{2}$International Centre for Radio Astronomy Research, Curtin University, GPO Box U1987, Perth WA 6845, Australia \\[0.3em]
  \small Correspondence: K. S. Servis, \texttt{knservis@gmail.com} \\
  \small (Received 18 October 2025; revised 27 January 2026; accepted 23 February 2026) \\
  \small Published in \textit{Publications of the Astronomical Society of Australia},
         \href{https://doi.org/10.1017/pasa.2026.10168}{doi:10.1017/pasa.2026.10168}
}}

\date{}

\maketitle

\begin{abstract}
Estimating the meteoroid flux density at centimetre to metre sizes is notoriously difficult.
Yet it is an important endeavour, as these sizes represent the transition between small meteoroids that pose a risk to spacecraft, and the Near-Earth Objects that are relevant for planetary defense.

We present a novel automated methodology for debiasing meteor observations from multi-camera networks, applied to data from the Desert Fireball Network (DFN). Our approach utilizes the Hierarchical Equal Area isoLatitude Pixelisation (HEALPix) framework to partition the sky into equal-area pixels at 70 km altitude, enabling precise and convenient measurement of effective survey coverage and fireball counting  across the network.
We developed a comprehensive data processing pipeline that analyses millions of all-sky camera images to determine clear-sky conditions through automated star source detection and flux distribution analysis.

As a case study, we apply this methodology to observations of the 2015 Southern Taurid meteor shower, during which there was significant fireball activity.
Processing data from 33 cameras over a three-month period (October-December 2015),  we calculate an effective observation coverage of $1.58 \times 10^{12}$ km$^2$.h and identified 54 Southern Taurid fireballs from 141 validated detections.
Our results are consistent with the extrapolation of previous work done on the same meteor shower at smaller sizes, when we set a $\sim300$ kg.m$^{-3}$ mean meteoroid density, consistent with the cometary origin of the Taurid stream.

The HEALPix-based approach successfully automates what was previously a labor-intensive manual process, providing a scalable solution for accurate flux measurements from distributed camera networks; it is directly applicable to other meteor surveys.

\end{abstract}


\section{Introduction}\label{introduction}

The Desert Fireball Network (DFN) is a multi-camera network designed to
determine meteorite fall positions and pre-atmospheric orbits of
meteoroids. The network consists of autonomous observatories spread
across the Australian continent, capturing nightly observations of
fireballs \citep{howie_how_2017}.
  One of the aims of the network is to determine meteoroid flux density at 1 AU at centimetre to decimetre sizes.

The necessity of accurately characterizing the meteoroid flux is
underscored by incidents such as the unexpectedly large meteoroid impact
on the James Webb Space Telescope (JWST) mirror
\citep{moorhead_library_2023}. Improved predictive models are
required to assess potential risks, especially from meteor streams like
the Taurids, known to contain significant numbers of larger objects
within resonant debris streams
\citep{devillepoix_taurid_2021,spurny_discovery_2017}.
Higher-order variations, such as latitudinal, daily, and seasonal differences, must
also be considered to achieve reliable predictions
\citep{robertson_latitude_2021,ozerov_goes_2024}.

At millimetre size, the large number of recorded impacts by a single  ground-based meteor camera readily enables population numbers studies \citep{blaauw_optical_2016,vida_computing_2022}.
At metre sizes and beyond, \citet{brown_flux_2002,brown_500-kiloton_2013} have used orbital sensors with global coverage.
These studies suffer from low number statistics, but are somewhat easier to de-bias (as long as the system sensitivity is well understood, transient effects like cloud cover do not get in the way of detecting bolides from orbit).

In the centimetre to decimetre range, the number of impacts detected by a ground-based network drops rapidly owing to the steep population index; de-biasing observations requires careful analysis \citep{halliday_detailed_1996}.
An accurate measure of the sky and the amount of time that the network was
observing. The DFN consists of approximately 30 cameras, which may or
may not be observing given the operational conditions at the time of
observation, including weather, time of night, moon altitude, equipment
failures, or misconfiguration. With a network of that size, controlling
for all these different factors that may affect observation is
challenging. Additionally, combining the effective observed volume is
complex \citep{devillepoix_global_2020}.

The only significant attempt in tackling the centimetre to decimetre range is the work of \citet{halliday_detailed_1996} using the Canadian camera network of the Meteor Observation and Recovery Project (MORP).
MORP consisted of 12 stations, with five rectilinear cameras at each, tiling the visible sky.
The network observed mass ranges from 1 g to hundreds of kilograms \citep{halliday_detailed_1996}.
Observation frames were divided into quadrants between
elevations of 8-20°, and 20-58°, with each quadrant rated as clear or cloudy. The sky at 70 km altitude was divided into cells, and
visibility from multiple cameras was assessed to arrive at a total
observation area and clear observation time. Meteors that did not pass
through these defined cells were discarded.

In this study, we present the methodology we have developed for estimating the influx rate of meteors detected by the DFN.
As a case study, we apply this method to observations collected
during the period when the Southern Taurids meteor shower showed
increased activity at fireball sizes, as documented by \citep{devillepoix_taurid_2021}.
Given DFN's capability to estimate mass kinematically
\citep{gritsevich_meteor_2008,sansom_novel_2015}, an accurate
estimation of influx rate can significantly enhance our understanding of
meteoroid dynamics and associated risks.

\citet{ehlert_measuring_2020} discusses several methods for debiasing meteor flux measurements in optical surveys, particularly those relating to the NASA All-Sky Fireball Network.
A notable claim from that paper is that, despite advances in analysis and automation, no fully automated, universally robust method has been found to reliably identify
clear-sky periods from all-sky camera data alone.

\citet{campbell-brown_optical_2016} addresses the challenges of combining observations from different meteor cameras and the necessary steps to debias result flux measurements for scientific comparability.
The methods used in that study include standardisations of flux measurements using a common reference limiting magnitude. They achieved inter-camera flux consistency through manual calibration and magnitude normalization, our method generalizes this approach to millions of all-sky images, automatically quantifying clear-sky conditions and network coverage. This enables unbiased flux estimates at centimetre to metre scales, where manual normalisation becomes infeasible.

\section{Data}\label{data}

To estimate the meteoroid impact flux density on Earth, two things are needed: a surveying system that accurately detects fireballs, and a way to measure how much area and time the system has effectively surveyed for.

The surveying system consists of the DFN observatories' main imaging system: Nikon D800/D800E/D810 associated with a Samyang 8mm f/3.5 UMC Fish-eye CS II, with a liquid crystal shutter.
Capture mode is 25-second-long exposures every 30 seconds, ISO 3200 or 6400, with the lens operated at f/4.
The liquid crystal shutter is operated at 10 Hz (see \citet{howie_submillisecond_2017} for details).
The system captures images continuously from nautical twilight (Sun altitude $<-6$ degrees), unless it detects that the conditions are too cloudy or if there is some technical issue preventing observations
\citep{howie_how_2017}.
The data it produces are 14 bit raw images in the
camera's native lossless compression file format (Nikon NEF).

The 10 Hz liquid crystal shutter provides temporal modulation of fireball trails, enabling flight direction and timing information to be derived during subsequent trajectory reduction, rather than from image timestamps alone. Observation start and end times rounded to minutes or 15-minute intervals, as reported in the survey coverage analysis, reflect operational selection boundaries for data processing, not the intrinsic timing precision of individual fireball detections. Individual fireball timestamps may be reported with sub-second resolution as they originate from the trajectory reduction pipeline. The flux density calculations presented here do not rely on millisecond-level absolute timing accuracy, but rather on the completeness and consistency of the survey coverage metrics.

\subsection{Fireball detection and reduction}\label{sec:fireballs}

The software that detects fireballs is described by
\citet{towner_fireball_2020}; this task is run on board the observatories.
The results are logged and transferred to the central server, the fireball candidates then get reviewed by a human, and if confirmed as fireballs they are systematically ingested.
Automated routines then retrieve the raw image data, as well as images suitable for astrometric calibration.
Calibration images are automatically solved following \citet{devillepoix_dingle_2018}.
The fireball track gets decoded and picked by a human, and the picked pixel coordinates are converted to sky coordinates.
If a fireball has been detected by more than one camera, it gets triangulated using the least-squares method of \citet{borovicka_comparison_1990}.
Speed is estimated using the Kalman filtering method of \citet{sansom_novel_2015}.
The pre-atmospheric orbit of the meteoroid is calculated following the code of \citet{jansen-sturgeon_dynamic_2020}, with Monte Carlo runs establishing the uncertainty based on the uncertainty on speed and direction.

The mass at the beginning of luminous flight (\(m_0\)) is estimated using the dynamic trajectory analysis method of \cite{gritsevich_estimating_2008} for large, decelerating meteoroids, and the method of Stevenson et al (2025; in prep.) for small meteoroids showing minimal deceleration. Both methods proceed by first deriving a ballistic coefficient (\(\alpha\)) from a curve fit to raw altitude/velocity data. The ballistic coefficient reflects the ratio between the drag and weight forces acting upon a meteoroid during atmospheric flight. From this value, entry mass can be estimated using the following formula, which assumes uniform ablation, spherical morphology, and a consistent bulk density \citep{sansom_determining_2019}.
\begin{equation} \label{eq:entry_mass}
\ m_0 = \left[ \frac{c_d A_0 \rho_0 h_0}{2 \alpha \rho_m^{2/3} sin\gamma} \right]^3
\end{equation}
\(c_d\) and \(A_0\) refer to the drag and shape change coefficients of a meteoroid. Following a commonly adopted simplifying assumption in dynamic meteoroid modelling \citep{gritsevich_estimating_2008, gritsevich_meteor_2008}, the product \(c_d A_0\) is taken to be 1 in the present analysis. \(\rho_0\) and \(h_0\) are respectively the critical density and altitude of the exponential atmosphere, as defined by \cite{gritsevich_meteor_2008}. Finally, \(\rho_m\) and \(\gamma\) are the bulk density and entry angle of the meteoroid.

The results in the present work use the same set of fireballs from the Southern Taurids stream as \citet{devillepoix_taurid_2021}.

\subsection{Data used to determine effective observing time}

The area and time the system has effectively surveyed are not
automatically generated as a data product. As images are received by the
on-board computer, a rough assessment using star counts is performed for
operational purposes, to decide whether the system should continue
observing (and to save shutter actuations). However, this system has a
very high threshold and only triggers when the sky is completely
overcast, to avoid missing very bright fireballs that can penetrate through thin cloud layers.

A decision was made early on to keep all the data captured by the
cameras, even images that do not contain fireballs.
There were multiple
motivating factors for storing all that data:
\begin{itemize}
\item  being able to go back into the archive to find fireballs that had too low signal to noise to
be detected, but could be useful to constrain the trajectory of a
fireball detected by other observatories \citep{shober_arpu_2022}.
\item  be able to later mine the data for other transients, notably astrophysical transients
\citep{ligo_scientific_collaboration_and_virgo_collaboration_properties_2019, kann_grandma_2023}.
\item  keeping all the data provided some contingency as well as validation
data when the fireball detection software of
\citet{towner_fireball_2020} was being commissioned.
\item  enabling the clear-sky survey, subject of the present work.
\end{itemize}

Keeping the data was enabled on the observatories side by the
availability of appropriate temporary storage (terabyte-scale hard drive
disks), and data centre space available to use for the program (petabyte on tape).
Maintenance trips include among their tasks the replacement of hard drives for empty ones.
The drives are then returned to the
Perth campus of Curtin University, and are uploaded via a high-speed link to the nearby Pawsey supercomputing research centre's systems.
The
permanent storage system that was provided by Pawsey for that use is a tape-backed media database called Mediaflux that is unsuitable for mass
parallel access to the images and logs.
Originally two 4TB hard drives were used on-board the cameras, meaning that the raw files would
fill up the disk space after about 4 months of clear weather. Thanks to
larger capacity drives becoming available and an added slot in the
observatories, this capacity was extended to three 10TB drives, which
gave each observatory enough space to store \textasciitilde1.5 years of raw data.
We later implemented another strategy to increase how long the systems could run autonomously: once the disks were near full (\textgreater90\%), a data compression task was run on the oldest data
that did not contain fireballs. The compression turns the 14 bit raw
\textasciitilde45MB images into full-resolution 8-bit colour JPEG files,
a \textasciitilde10:1 compression factor, this is enough to ensure that
hard disk space is not a limiting factor for continued observations
(other parts such as the mechanical shutter would fail first).

We use this compressed data product to calculate the effective clear-sky
time-area product. For the purpose of determining clear sky conditions,
it was found that the lossy compressed JPEG images are nearly as good as
the raw data product. This means that we only need to work with a more
manageable \textasciitilde300TB dataset, rather than the full 3PB
archive. In some cases that was the only option as when the
observatories run out of disk space, such as when a maintenance cycle
needed to be extended so the disks were not replaced with empty ones, only the compressed version is kept.

The network has been in continuous operation since mid-2014.
For this project we extracted a full year of data (2015) from the tapes.
For validation purposes we focus on the period from 2015-10-01 to 2015-12-31, as
images from that period had been fully examined manually by \citet{devillepoix_taurid_2021}, providing a good basis for
validation of the methods used.
This period was chosen due to its association with the Southern Taurid meteor shower \citep{devillepoix_taurid_2021,egal_proposed_2022}.
Additionally, observations in that period all have associated orbit and derivative
information (e.g., meteor shower, radiant). The calculated radiants for
the period of activity in 2015 (blue to red colour is the solar
longitude) can be seen in Figure~\ref{fig-radiants}. It is worth noting
that these dates were used for final processing but the results for the size--frequency distribution (SFD) are related to only part of that period as that was the actual period of activity.
Specifically, the Southern Taurid period of activity
was between 2015-10-27 and 2015-11-17 (solar longitudes 213° and 234°, respectively).
Over that period, 27 unique cameras recorded images. The total hours of observation anywhere in the network was 245.75 over that period. If all cameras were observing over that whole period that would give a total image count of 796230. Due to a combination of optimisations (for sun altitude calculated locally, being different across the network) and the occasional malfunction the actual number of images recorded over that period was 385284, which is still a very large number to sort manually and that only corresponds to a single event.

As detailed later in Sec. \ref{clear-sky-survey-pipeline}, when evaluating how clear an image is, the astrometric mapping of the camera must be known.
To achieve this at scale, for each image that needs to be evaluated for clear sky conditions, we use the astrometric solution nearest in time already calculated for this particular camera as part of the fireball data processing pipeline (Sec. \ref{sec:fireballs}).

\section{Methods: Clear sky survey pipeline}\label{sec-methods}

The end product of the clear sky survey pipeline is a measure of how much Earth area is being monitored for meteoroid impacts over time.

\subsection{Discretisation of the area-time monitored for fireballs}

The millions of images to process, taken over a long period by a network of irregularly spaced sensors, calls for discretising the time and area we are working with.
The time discretisation choice is straightforward: the cameras take picture synchronously over the entire network every 30 seconds.

However, the discretisation method for the area covered is not evident.
The on-sky effective field of view of the all-sky sensors varies because some are affected by obstructions (trees, antennas, buildings, etc.).
It is also not rare for a sensor to be partially cloudy.
Ideally one would like to be able to retain significant granularity when assessing the effective clear observable sky of each sensor.
Furthermore the monitored areas in the atmosphere need to be easily aggregated between neighbouring camera systems in the network, in a scalable way.

The Hierarchical Equal Area isoLatitude Pixelisation (HEALPix) framework developed by
\citet{gorski_healpix_2005} provides a way to partition the sky into equal-area regions,
which is particularly useful for a multi-camera all-sky survey like the one described here.
In simple terms, HEALPix provides a mapping from spherical coordinates to discrete sky regions (called HEALPix elements).
The spherical coordinates can be geographic (latitude, longitude) or celestial (right ascension, declination).
Because HEALPix elements have equal area on the celestial sphere, they are suitable for aggregating observations from different camera locations.
The mathematical method for mapping spherical coordinates to HEALPix elements is described in detail by \citet{gorski_healpix_2005},
and the implementation is provided by \texttt{AstroPy} \citep{robitaille_astropy_2013}.

\begin{figure*}[ht!]
\includegraphics[width=1.0\textwidth]{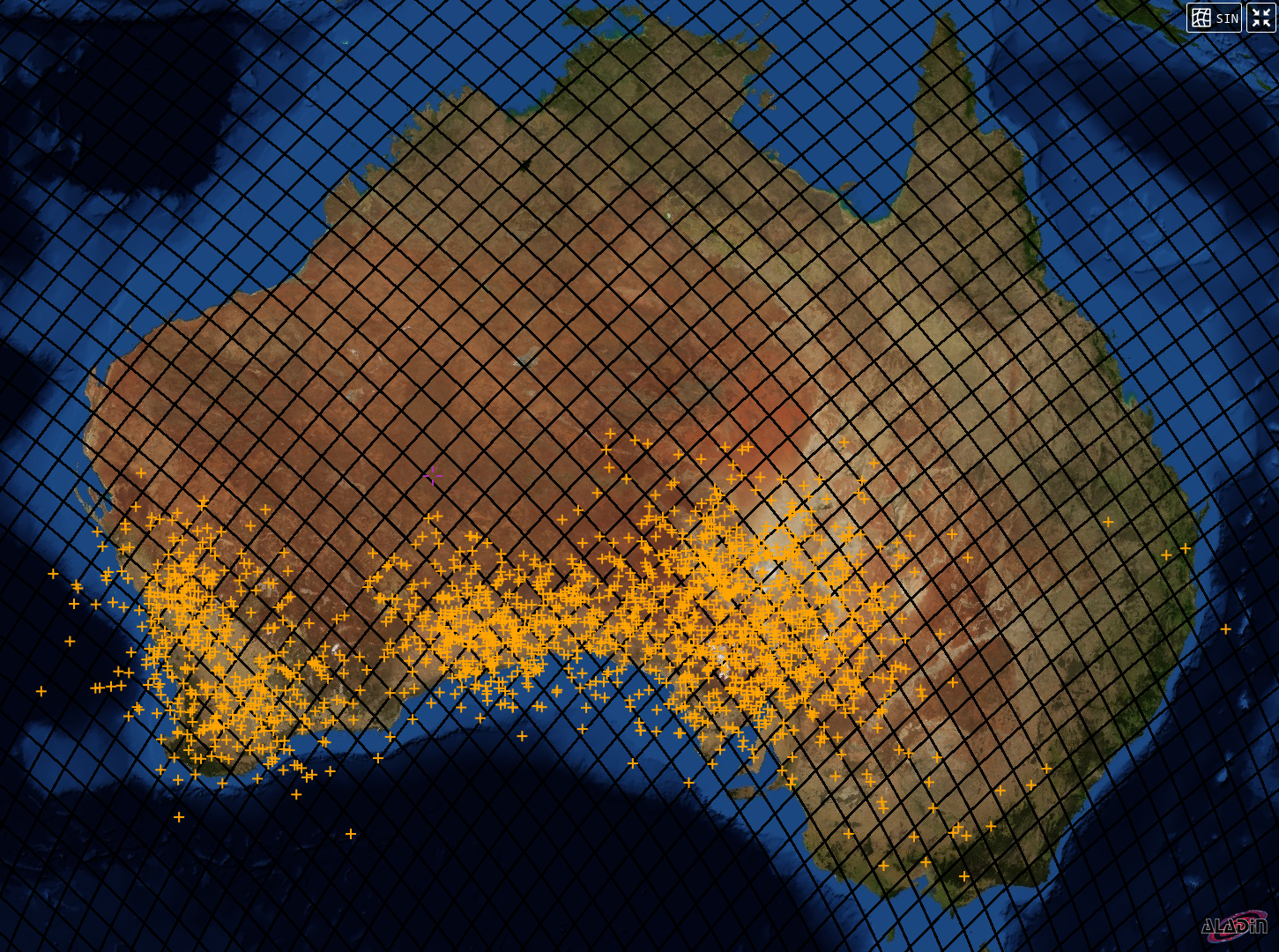}
\caption{\label{healpix-example}
Black grid: HEALPix grid at order 6 over Australia (individual HEALPix elements shown as rhomboid cells with $\sim$103\,km side length, $10,630\,\text{km}^2$ in area). Orange +: impact locations of all meteoroids detected by the DFN.}
\end{figure*}%

\subsection{Processing steps}\label{clear-sky-survey-pipeline}

\begin{figure*}[ht!]
\includegraphics[width=0.95\textwidth,keepaspectratio]{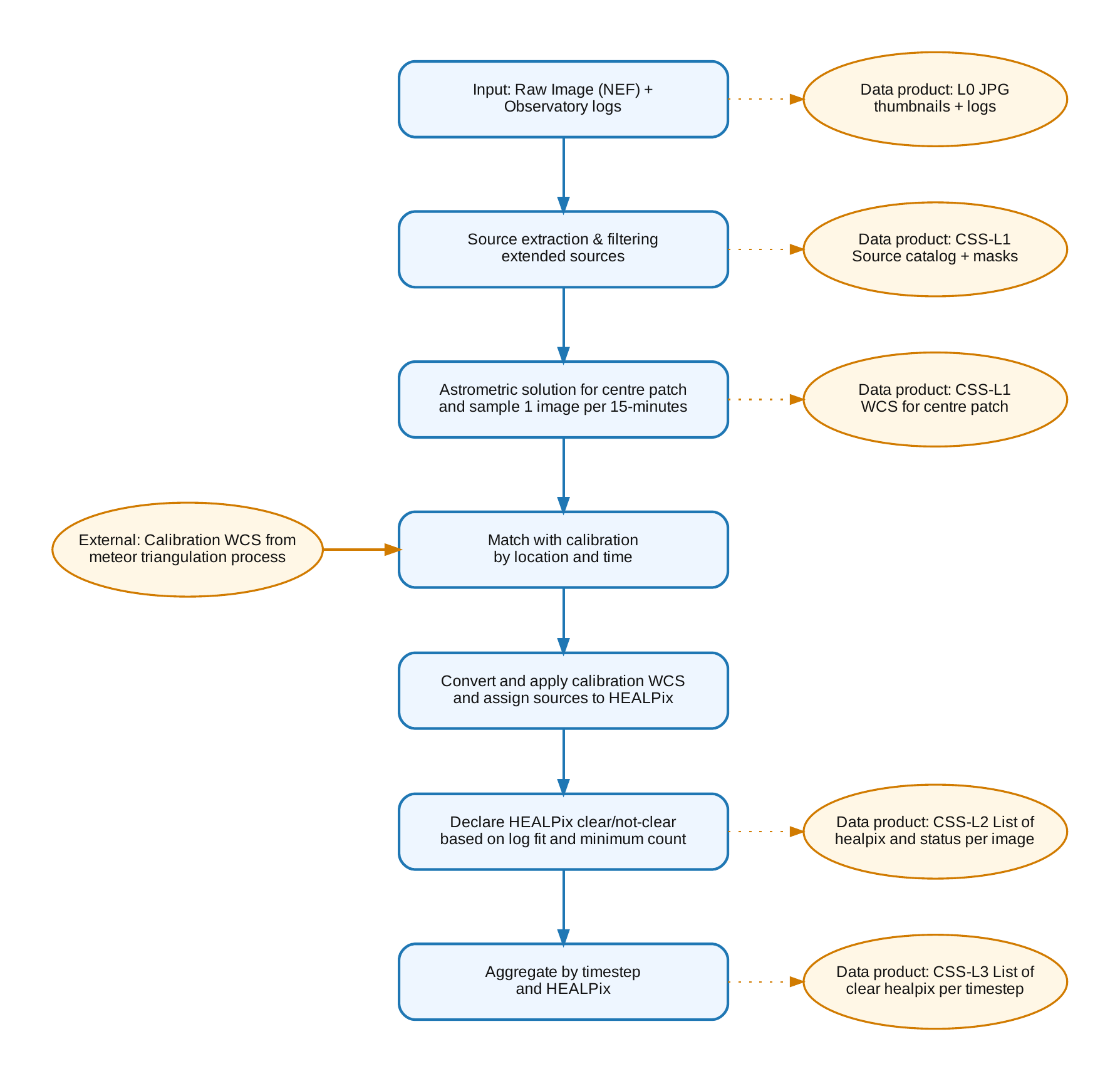}
\caption{\label{processing-flowchart}
Processing steps to arrive at a clear sky data product. Blue rectangles: high-level processing tasks. Orange circles: Data products.}
\end{figure*}%

The data processing pipeline we have developed is organised in several steps that each generate intermediate data products (Figure~\ref{processing-flowchart}). These products follow a hierarchical structure: \textbf{Level 0 (L0)} consists of raw compressed images and engineering logs; \textbf{Level 1 (L1)} provides per-image star catalogues derived from source detection; \textbf{Level 2 (L2)} assesses clear-sky status for individual equal-area sky regions from each camera; and \textbf{Level 3 (L3)} aggregates these measurements across the entire network to produce the final survey product.

These intermediate data products are organised using two data storage solutions.
Firstly, large files are stored on collections (called buckets) on the Acacia object store system provided by Pawsey (a Ceph-based \citet{weil_ceph_2006} S3 compatible object store); we refer to this system below as the "object store".
Secondly, a NoSQL document database (using MongoDB) keeps track of the existence and locations of all the files in the object store and their metadata; we refer to this system as the "database".

The database also keeps track of the overall progress on the data processing steps.

\paragraph{Level 0: JPG images and engineering logs}
The raw images and engineering logs on tape for all the years of operation of DFN amount to \textasciitilde2.5PB of data.

In batches, these files are retrieved from the tape store, and staged to a temporary fast storage area at the Pawsey Supercomputing Centre.
The raw (Nikon NEF) file format is converted to a full resolution JPEG.
These compressed images (~10:1 compared to raw) and engineering logs form the L0 data product.
They are organised in buckets on the object store.

The Mongo database keeps track of the existence and location (URL) in the object store of all the files ingested and their metadata, including images and engineering logs which are logs produced by the observatory for the observing session and are stored next to the images for the session in the object store.
The metadata stored in the Mongo database includes the time and location of capture of the data for this level.

\paragraph{Level 1: Per image light source catalogue}
The next level of data product (Level 1) is produced by performing the following steps:
\begin{enumerate}
    \item Select one image in each 15 minute intervals (we assume observing conditions do not change significantly in 15 minutes).
    \item Image preprocessing to mask out contiguous
regions that are not showing stars. These correspond to the Moon and
surrounding sky where stars cannot be seen and ground objects, such as
trees and masts.
    \item Extracting the point sources (stars) pixel coordinates and brightness using the Python \textit{Source Extractor} library (SEP; \citet{barbary_sep_2016}).
    \item Astrometrically solve using astrometry.net \citet{lang_astrometrynet_2010} the centre patch of the image. Only the centre patch is used as there are significant lens distortions away from the centre. This is used for verification and for cataloguing.
    \item Converting the stars pixel coordinates to equatorial sky coordinates. For this step we re-use astrometric solutions that are generated automatically for doing fine astrometry on detected bolides and convert the solution to the current time. That solution is computationally expensive and takes into account the significant lens distortions. To facilitate this process we ingested the transformation parameters for all astrometrically solved images into the Mongo database (typically one solution per camera system every couple of days). For each clear-sky survey image analysed we queried the astrometric solution closest in time for the same observing system and translated the solution to the current sidereal time.
\end{enumerate}

\begin{figure*}[ht!]

\includegraphics[width=1.0\textwidth]{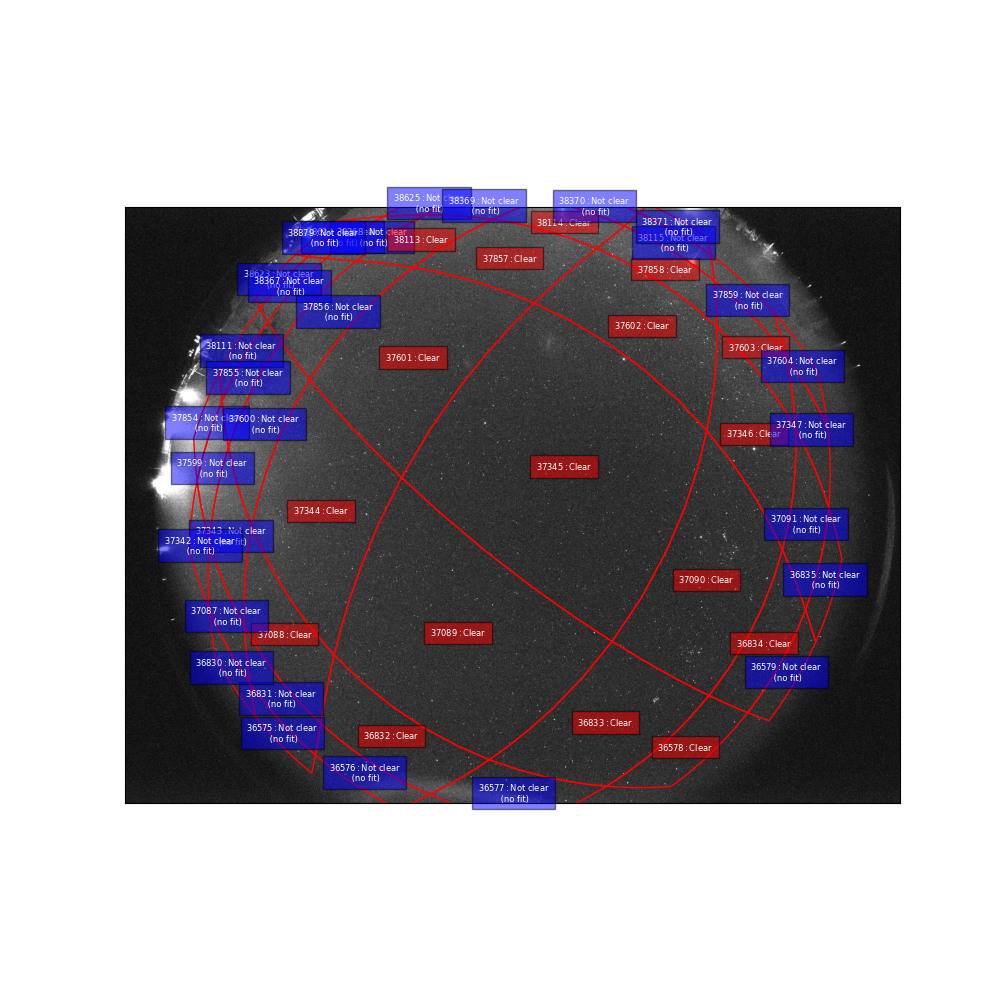}

\caption{\label{fig-count-by-healpix}Clear HEALPix example. This figure illustrates the distribution of clear-sky conditions across HEALPix elements for a single observation window. Each colored cell represents a HEALPix element at 70~km altitude that was classified as clear based on the flux distribution of detected sources.}

\end{figure*}

\begin{figure*}[ht!]

\includegraphics[width=1.0\textwidth]{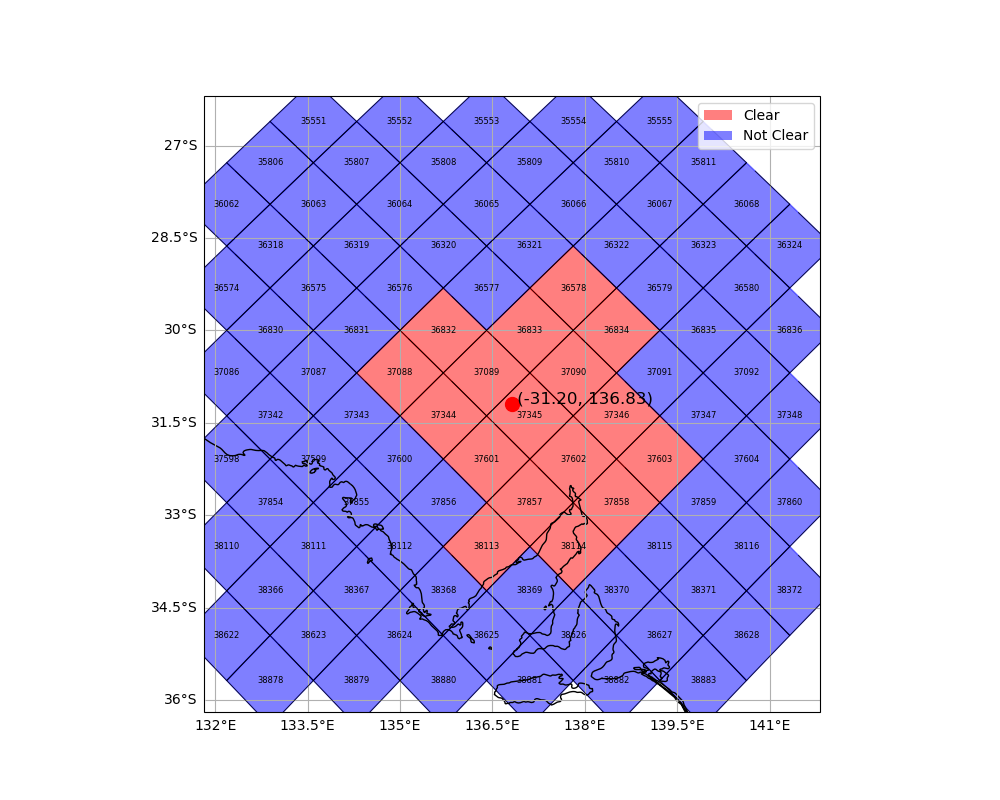}

\caption{\label{fig-single-camera-clear}Single camera clear HEALPix example. This panel shows the clear-sky HEALPix elements as observed by a single DFN camera during a 15-minute window. The highlighted HEALPix elements indicate areas where the flux distribution of detected sources met the criteria for clear-sky classification (logarithmic fit $R^2 > 0.75$, at least 10 sources).}

\end{figure*}

\paragraph{Level 2: Per-camera clear-sky status}
The L2 data product consists of calculating the pixel
boundaries on the image for the 70km atmospheric shell HEALPix (see
Section~\ref{sec-healpix} for details) as they would appear from the
location of the camera. The sources are then matched to the
corresponding HEALPix element number and if the sources fit a logarithmic
function down to a threshold and there are enough sources visible then
the HEALPix element is considered clear for that location on that 15 minute
window. 
Specifically, for this step, the sources for that window are
sorted according to the measured flux. If a logarithmic fit (log10) to
the flux distribution yields a coefficient of determination (\(R^2\)),
calculated using \texttt{sklearn.metrics.r2\_score}, greater than 0.75,
and there are at least 10 sources detected within that HEALPix element,
the region is classified as ``clear.'' That is then recorded in a
``clear HEALPix element'' entry for each image and record in a file next to the
image on the Mongo database (see figures \ref{fig-count-by-healpix} and \ref{fig-single-camera-clear} for illustrative examples) 

This clear-sky classification is based on the shape and goodness-of-fit of the flux distribution, not on a single limiting magnitude threshold. A fixed limiting magnitude is not assumed, as camera sensitivity and atmospheric conditions vary between observing sessions. Thin or patchy clouds may still permit detection of bright stars, but typically distort the expected logarithmic flux distribution, causing the HEALPix element to be classified as not clear. The minimum source count requirement (10 sources per HEALPix element) helps avoid misclassification under marginal conditions where few detections could spuriously fit the distribution criterion. By design, this conservative approach prioritizes accurate effective observing time estimation over maximizing apparent coverage, ensuring that the survey area reflects conditions suitable for reliable fireball detection and flux analysis.

\paragraph{Level 3: Network-wide clear-sky clear}
The L3 data product consists of aggregating the clear HEALPix element classifications 
across all cameras and time steps to create a final data product of multi-camera HEALPix element conditions 
over the entire survey area. 
This data product provides the effective collection area for the phenomena observed and is used for debiasing the observations.

\begin{figure*}[ht!]

\includegraphics[width=1.0\textwidth]{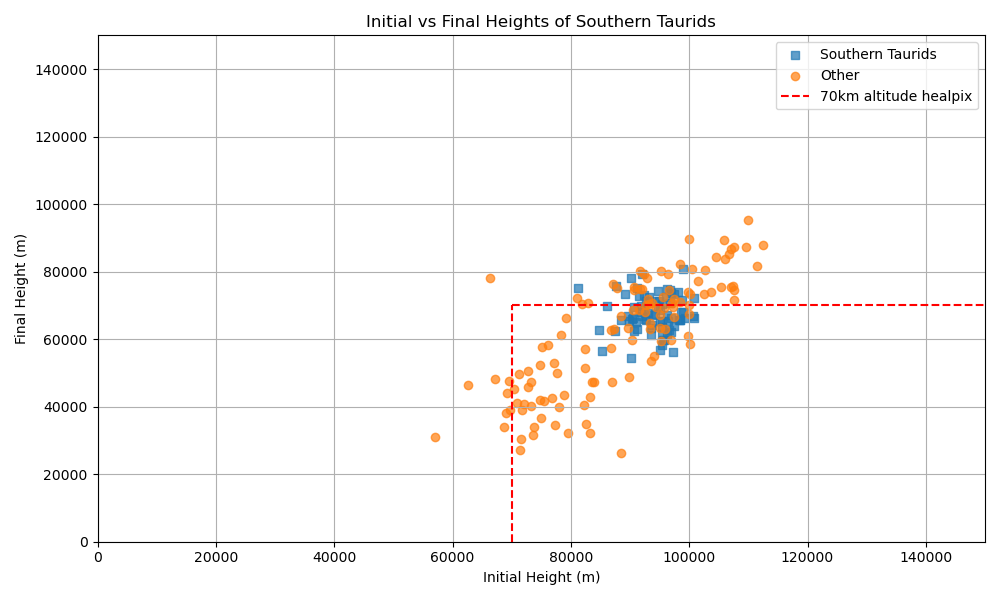}

\caption{\label{fig-70km-selection}Start vs end height distribution of detected 
meteors. This figure shows the distribution of start vs end height of the light path,
 highlighting the 70 km altitude shell used in the DFN clear sky survey.}

\end{figure*}%

\begin{figure*}[ht!]

\includegraphics[width=1.0\textwidth]{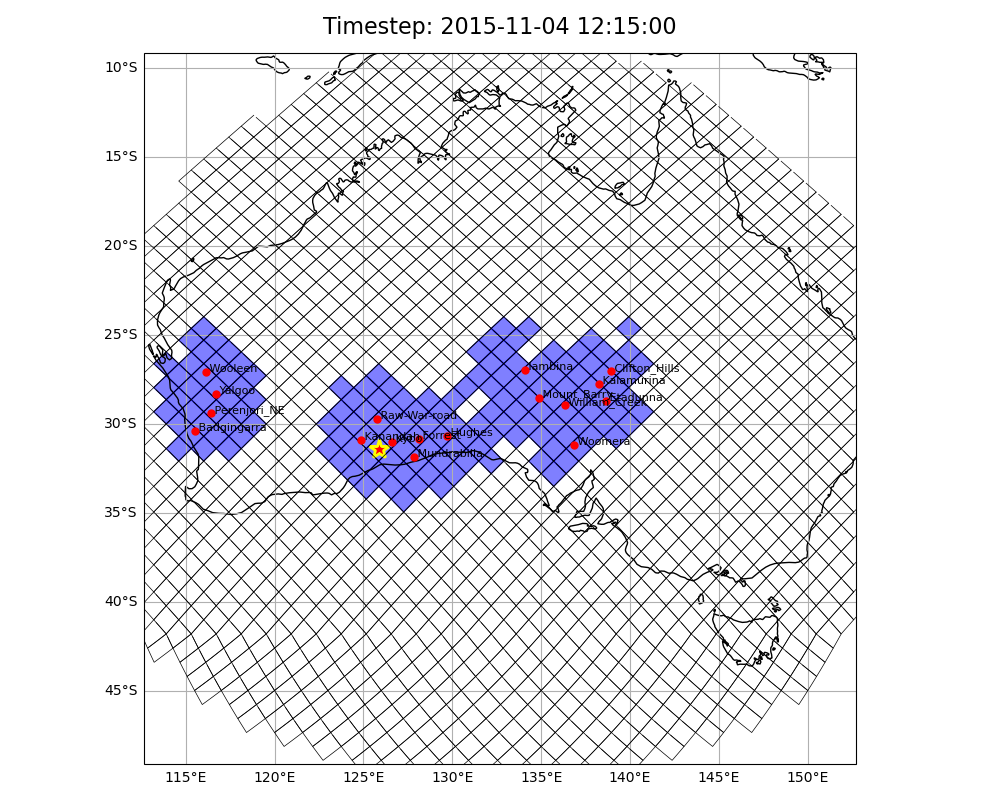}

\caption{\label{fig-single-timestep-example}Single time step with clear HEALPix elements and detected fireball. This figure presents a snapshot of the network's coverage at a specific time step, showing all HEALPix elements classified as clear by at least two cameras. The positions of the cameras and the HEALPix element containing a detected fireball are overlaid.}

\end{figure*}

All the clear HEALPix windows are collected by timestamp, including the
identifiers of the cameras that have been observing that particular
window at that time (see Figure~\ref{fig-single-timestep-example}).

\section{Results}\label{sec-results}

We process the data for last three months of 2015 (2015-10-01 14:00 to 2015-12-31 20:15 UTC).
This period fully encompasses the Southern Taurid activity period (first recorded fireball occurred on 2015-10-30T12:04:38 and the last on 2015-12-08T13:13:48).
The first section in this results section consists of a summary covering the three months including quality controls for the derived clear-sky data.
The subsequent section (\ref{meteoroid-flux-density}) aims to calculate the impact flux density for the 2015 apparition of the Southern Taurids meteoroid stream.

\subsection{Overall time-area coverage}\label{sec-area-coverage}
The clear-sky coverage for the last three months of 2015 is discretised using a 15 minute timestep.
During each timestep a camera is considered to be effectively observing if it has captured 30 images.
This condition ensures that the data is complete in  in the sense that the capture control system worked as configured.

Timesteps that we excluded were usually because they overlapped with the start or end of operation for the day.
Other causes for excluding these camera timesteps include occasional malfunctions that led to missed shots, and some systems being in test mode (e.g.~a camera's location identified as ``Test\_lab''). A list of the included cameras operation time can be found on Table~\ref{tbl-cameras}.

During the period of study, we assessed a total of 4213 timesteps of 15 minutes between dusk and dawn.
The total number of images for the period was 1713609 from a total of 33 unique cameras.
For all timesteps the cameras observing
were on average 12.54 and not observing 1.89 meaning that they were excluded from the survey for any one of the aforementioned reasons.

The location of the cameras as defined by their GPS coordinates is not
stable throughout the observation and drifts by a few metres over the
span of the survey, so they were assigned to the centre of their
corresponding box containing these coordinate points with a maximum distance of 100 metres from that centroid.
The reasons for such drift can be rounding errors and GPS inaccuracies but also movement/substitution of the camera by the maintenance team during deployment or maintenance.
Approximate locations of the cameras and
the width and length of the corresponding bounding box that contains all
the recorded GPS points is given in Table~\ref{tbl-unique-cameras}.

Each HEALPix has a specific number (e.g.~37089 was one of those as can be seen in Figure~\ref{fig-single-camera-clear}). That number is referring to that area in the sky and is the same between cameras, but viewed (if visible from that location) at a different angle. The number of individual HEALPix observed by any camera to be clear at any time in the
period was 217 while the number of clear HEALPix observed by more than camera
one was 157. That is because HEALPix at the edges of the network were never observed by more than one camera.

Each HEALPix was observed on average 1426 timesteps by a
single camera and 521 on average by more than one cameras. Observations for the
network start on average at 10:30 UTC and end on average at 20:30 UTC
with an average duration of just under 10 hours each day.

When we sum up all the HEALPix observed by at least two camera for that observation period we arrive at
an effective total observation of
$1.5817464 \times 10^{12} \; \mathrm{km^{2}\,h}$.
the full table detailing the amount of time each HEALPix is clear one at least two cameras is given as supplementary material in Table~\ref{tbl-healpix-sum-time}), and visualised in Figure~\ref{fig-obs-period-cover}.

\subsection{Meteoroid flux density}\label{meteoroid-flux-density}

In order to arrive at a debiased result for the flux density we select which events we can include in the count.
There are two filtering steps. Firstly we filter on events per
camera based on whether that camera was observing. For that we consider a camera to be observing, when there are exactly 30 images for that 15 minute timestep.
Secondly, at network level, we only include fireballs that have been observed through a window considered clear for the timestep during which the fireball happened.
Finally we select those observations that are attributable to the Southern Taurids.

For the period corresponding to the survey, the first fireball event is on
2015-10-01 14:00:00 and the last 2015-12-31 18:30:00.
The total number of events were 647 however only 585 are ``valid'' in the sense that only those are taken into account as they come form cameras that are
observing given the definition of observing in
Section~\ref{sec-area-coverage}, namely that were observed during a 15 minute timestep for which exactly 30 images were recorded. From the summary on
Figure~\ref{fig-summary-events} we observe no notable patterns of
rejection of events over the period.

For the second filtering step, as the trajectory of the fireball is
known, we use the latitude and longitude corresponding to the beginning
and the end of the bright flight and assign the fireball to the lowest
in altitude HEALPix that was clear for that timestep and observable by
more than one camera. Using that filter we keep 141 and reject 56
fireballs. In 115 observations the lowest HEALPix coincides with the
70km altitude HEALPix and in 26 it does not. As can be seen from
Figure~\ref{fig-70km-selection} where height of the shell is shown in
comparison to the dataset this is expected.

From the calculated radiants for the selected period (as already
calculated by \citet{devillepoix_taurid_2021}) we found 54 fireballs
that are part of the Southern Taurids and 87 other fireballs, and the
period for those Southern Taurid observations between
2015-10-30T12:04:38.626 and 2015-12-08T13:13:48.446. As described in
that study the meteoroid's heliocentric orbit is determined by
backtracking the meteoroid from the start of its luminous flight out of
Earth's gravitational influence to the Hill sphere, using Monte Carlo
methods to estimate uncertainties.

The dynamic mass, assuming a spherical cross-section for the meteoroids
using the calculation described by \citet{gritsevich_meteor_2008},
is shown in Figure~\ref{fig-masses}. 
For consistency with prior DFN analyses \citep{sansom_novel_2015}, an ablation coefficient of $4*10^{-8}$ kg/J is adopted. A complete description of the mass calculation can
be found in \citet{sansom_novel_2015}.

By using the calculated observation area rate for the activity period of the Southern Taurids (
$2.2658823 \times 10^{16} \; \mathrm{\frac{m^{2}}{yr}}$), 
we can arrive at the normalised influx rate table
Table~\ref{tbl-influx-rates-mass}.

\begin{figure*}[ht!]

\includegraphics[width=1.0\textwidth]{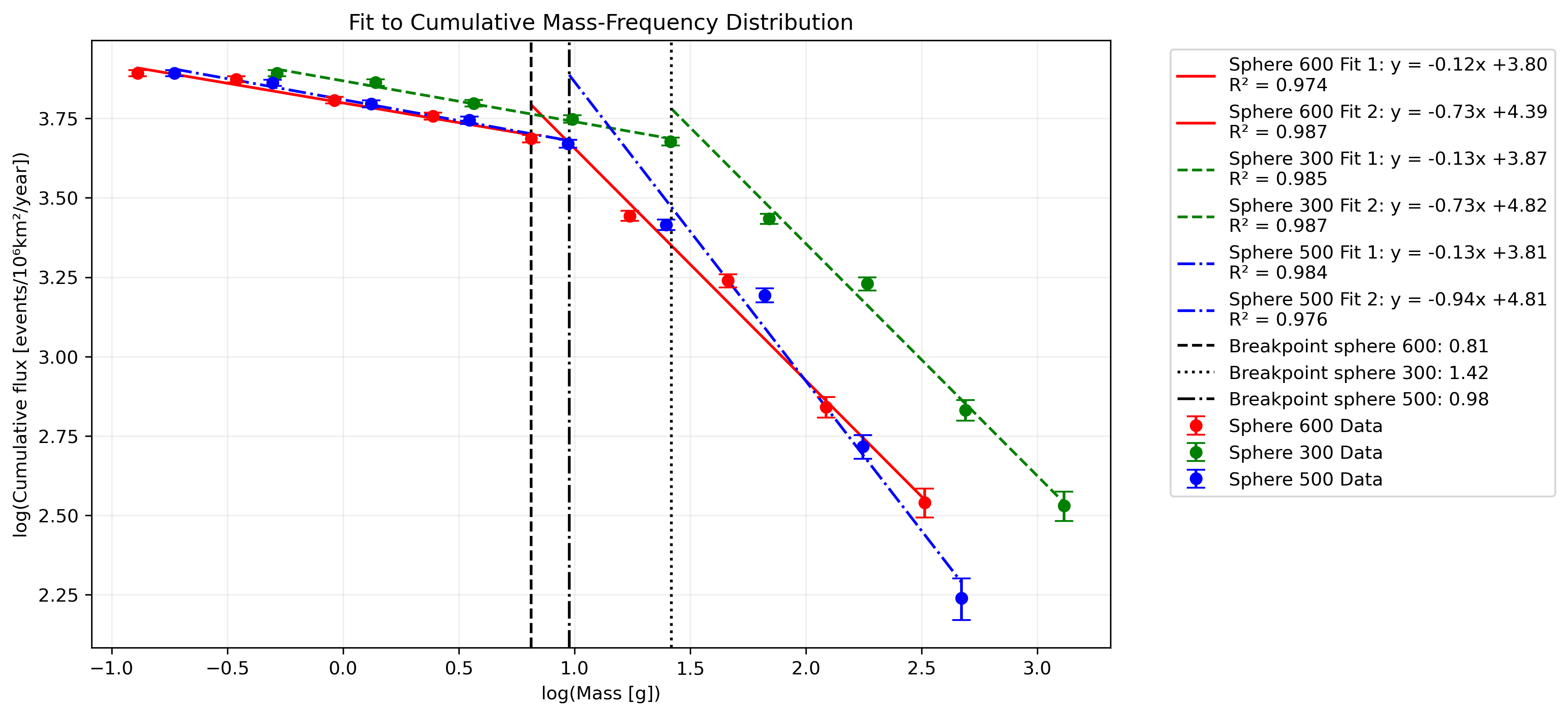}

\caption{\label{fig-log-fit-sfd}Fit to Cumulative Mass-Frequency
Distribution. This plot shows the cumulative size-frequency distribution
(SFD) of meteoroids detected by the DFN, with a fitted power-law model.
The presence of a break around 1 gram is highlighted.}

\end{figure*}

We can then calculate the cumulative size-frequency distribution, which
is presented in Figure~\ref{fig-log-fit-sfd} alongside the linear fit
for a spherical models for different density assumptions. The presence
of a discontinuity around the 1 gram point has been documented
previously by \citet{halliday_detailed_1996}. This discontinuity
represents a change in the distribution pattern, indicating that
particles or fragments below and above this mass threshold may have
different formation or fragmentation histories. The break or shift
suggests a physical or mechanical process that affects the population of
particles differently on either side of 1 gram, which could be due to
changes in fragmentation dynamics, aerodynamic sorting, or sampling
biases. In the case of the DFN part of this discrepancy will be
due to selection bias because the network is less sensitive to smaller
meteors.

The above fit corresponds to \(log N = a M_i + b\) with:

\begin{enumerate}
\def\labelenumi{\arabic{enumi}.}
\item
  a = -0.13 and b = 3.87 (mass index s = 1.13) for the first segment up
  to mass 1.42 and then
\item
  a = -0.73 and b = 4.82 (mass index s = 1.73)
\end{enumerate}

\section{Discussion}\label{discussion}

\begin{figure*}[ht!]

\includegraphics[width=1.0\textwidth]{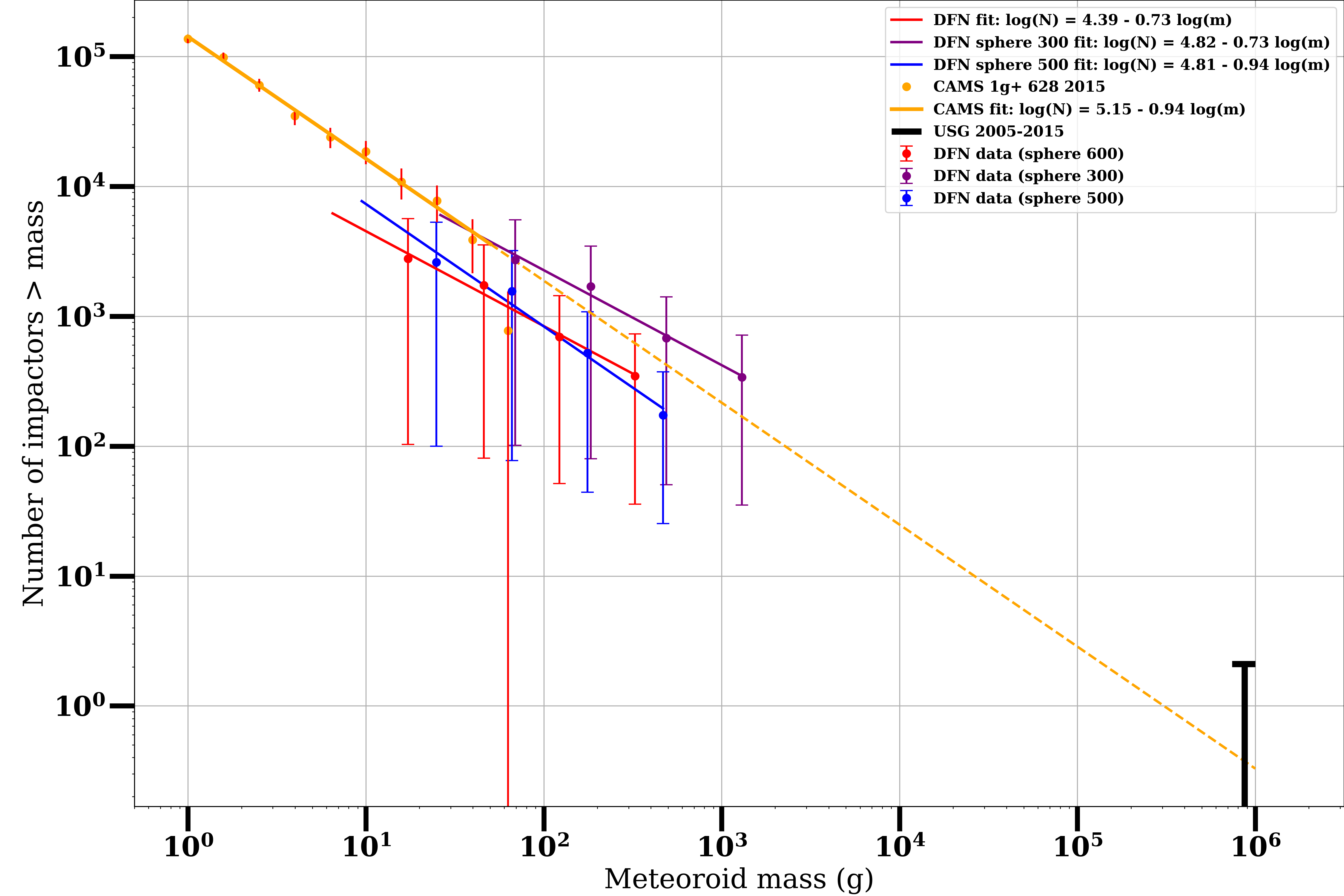}

\caption{\label{fig-sfd-with-cams}Comparison with CAMS results. This
figure compares the DFN cumulative mass-frequency distribution with
results from the CAMS survey, plotted for three different assumed
meteoroid densities (300, 500, and 600 kg/m³). The comparison
demonstrates the effect of density assumptions on mass estimates and
highlights the continuity and differences between the two surveys'
results.}

\end{figure*}

Compared to CAMS results presented by \citet{devillepoix_taurid_2021}
the results seem to complete the picture for that stream well extending
to lower masses. In Figure~\ref{fig-sfd-with-cams} the result is plotted
for 3 different assumed densities, namely 300, 500 and 600 \(kg/m^3\).
The resulting plot seems to form a better continuity when using a
\(300 kg/m^3\) assumption. This is a reasonable assumption based on the
cometary origins of this stream \citep{devillepoix_taurid_2021} as
these have been estimated to have densities around that range
\citep{ceplecha_earths_1988}. This does not mean that this is the
actual density of the meteoroids, as the calculation of the mass depends
on multiple assumptions about shape and physical properties. Mass
estimation in the CAMS data is done using the photometric method,
whereas here we use the dynamic mass \citep{sansom_novel_2015}.

Given the difference of the calculation and debiasing methods in the two
surveys, the data considered in this study cannot be considered to
conclusively estimate density; however, such an estimate is possible if
additional data, such as the photometric mass or a simultaneous radar observation, were added.

The use of HEALPix for debiasing meteor network observations has proven
highly effective at automating time-area coverage calculations across
large datasets. By dividing the sky into equal area pixels and tracking
clear sky conditions for each pixel over time, we can accurately measure
the effective survey volume of the Desert Fireball Network.

The resulting time-area measurements provide a robust basis for
calculating true meteoroid flux rates. By knowing exactly when and where
the network had clear views of the sky from multiple cameras, we can
properly normalize meteor counts and derive unbiased population
statistics.

While developed for the DFN, this approach could be valuable for other
meteor camera networks seeking to debias their observations and
calculate accurate flux rates. The use of standardized HEALPix
coordinates makes the results comparable between different surveys.

The dates that are used here are comparable with
the period of high activity for the newly
discovered branch of the Taurid meteor complex described by \citet{spurny_discovery_2017}.

The fitted slope is somewhat unexpected on both segments as, for the
first segment it shows as unusually flat in comparison to
\citet{halliday_detailed_1996} but that can be explained due to
observation bias. The masses recorded there are below one gram and the
resulting light path is equally small. As a consequence both due to the
optical limits and the automated detection limits described in
\citet{towner_fireball_2020}, leads to systematic bias.

Using HEALPix in meteor observations to debias observations can be
useful for other surveys and it is the intention to release this
functionality as a Python library for other users.

The method presented was also useful for adjusting the flux for biased
streams in an automated way, projecting the collecting area to the
perpendicular to the radiant. A future improvement would be to use
HEALPix which are on the orbital apex of the Earth orbit and assign also
that HEALPix number to an observation. That way any correction in terms
of the radiant of a focused stream would be much simplified. It is likely
a useful tool to much more easily visualise the shape of cometary debris
field as the Earth travels through it. Additionally, anisotropies in the
influx can more readily be evaluated given the equal area nature of
HEALPix.

Interestingly, even though there is insufficient data to estimate
meteoroid density in this study, such an estimate is a clear possibility
if we add data that can be made available. Specifically adding
constraints on the mass by using photometric methods could provide
sufficient explanatory power. Adding such information requires additional
analysis and can form the basis of future work.

In summary, using equal-area pixels (HEALPix) with the origin at the
geographic centre, coupled with logarithmic fitting of source flux
distributions, provides a robust and automated method for calculating
collection area across large datasets from multiple observatories.
Additionally, using a shell centred on the apex of Earth's motion
enables more accurate correction for directional biases in meteoroid
influx, facilitating improved flux normalization and inter-network
comparisons.

\section{Conclusions}

This study leverages the HEALPix framework, an equal-area
pixelization method, to accurately estimate the portion of the sky that
is clear and effectively map the spatial distribution of meteors.
HEALPix's equal-area pixels facilitate uniform analysis and easy
aggregation of observations from multiple cameras, significantly
simplifying the process of calculating observed areas and times
\citep{gorski_healpix_2005}. This approach enables precise, unbiased
assessments of meteor distributions and enhances our ability to
characterize spatial-temporal variations in meteoroid activity, crucial
for improving flux estimation and predictive models.

The overall aim to provide an automated way to estimate clear-sky conditions in a way that can be used in other multi-camera networks is successfully demonstrated by comparing the mass size-frequency distribution derived using the debiased influx, to other studies.

\section*{Acknowledgments}

Parts of this work were developed during the employment of K. Servis by the Commonwealth Scientific and Industrial Research Organisation at Pawsey Supercomputing and Research Centre in Perth, Western Australia and Centre National de la Recherche Scientifique at Laboratoire d'Astrophysique de Marseille in Marseille, France. 

This research made use of Astropy \citep{robitaille_astropy_2013}, a community-developed core Python package for Astronomy.

\onecolumn

\appendix

\FloatBarrier

\section{Clear HEALPix estimation method}\label{sec-healpix}

The aim of the DFN clear sky survey is to provide a robust automated
estimate of the amount of time and area surveyed when DFN observations
took place \citep{bland_rate_2006} in millions of images.

The method we used to estimate the observed volume by measuring
observations along a shell 70km above MSL following the clear-sky survey of \citet{halliday_detailed_1996} and measuring how
many cameras can view, which part of the shell at any given time. The
choice of 70km altitude was made so that the shell is representative of
a boundary that we expect the light-path to cross for all the objects of
interest that will be included in the final count. An overview of where
that shell lies can be found in Figure~\ref{fig-70km-selection} and
initial and final height of the observed fireballs and for those specifically belonging to the Southern Taurids is given.

So that the estimates of observations can be used between cameras and
can be extrapolated at a larger scale, we used the Hierarchical Equal
Area isoLatitude Pixelisation (HEALPix) framework developed by
\citet{gorski_healpix_2005}. We translated the WGS84 geographical
coordinates of Earth at an altitude of 70\,km to HEALPix elements with NSIDE 64,
which divides the observed shell into equal-area regions of 10606.58\,km$^2$. Each HEALPix element
was transformed into an image polygon at the time and
location of each camera, and each point light source observed was
assigned to the corresponding HEALPix element. By ordering the
sources in terms of flux using a minimum of 10 sources, if the fluxes
fit a logarithmic function, we consider that the origin is
astronomical and the HEALPix element for that camera and time was declared
``clear'' (see Figure~\ref{fig-count-by-healpix} and Figure~\ref{fig-single-camera-clear}).

Subsequently, the HEALPix elements for a given 15 minute timestep were collected
together to arrive at the total observation area for that time step (see
Figure~\ref{fig-single-timestep-example}).

\FloatBarrier

\section{Additional Figures}\label{sec-additional-figures}

This appendix contains supplementary figures that provide additional context to the survey results.

\begin{figure}[htbp]
\centering
\includegraphics[width=\textwidth]{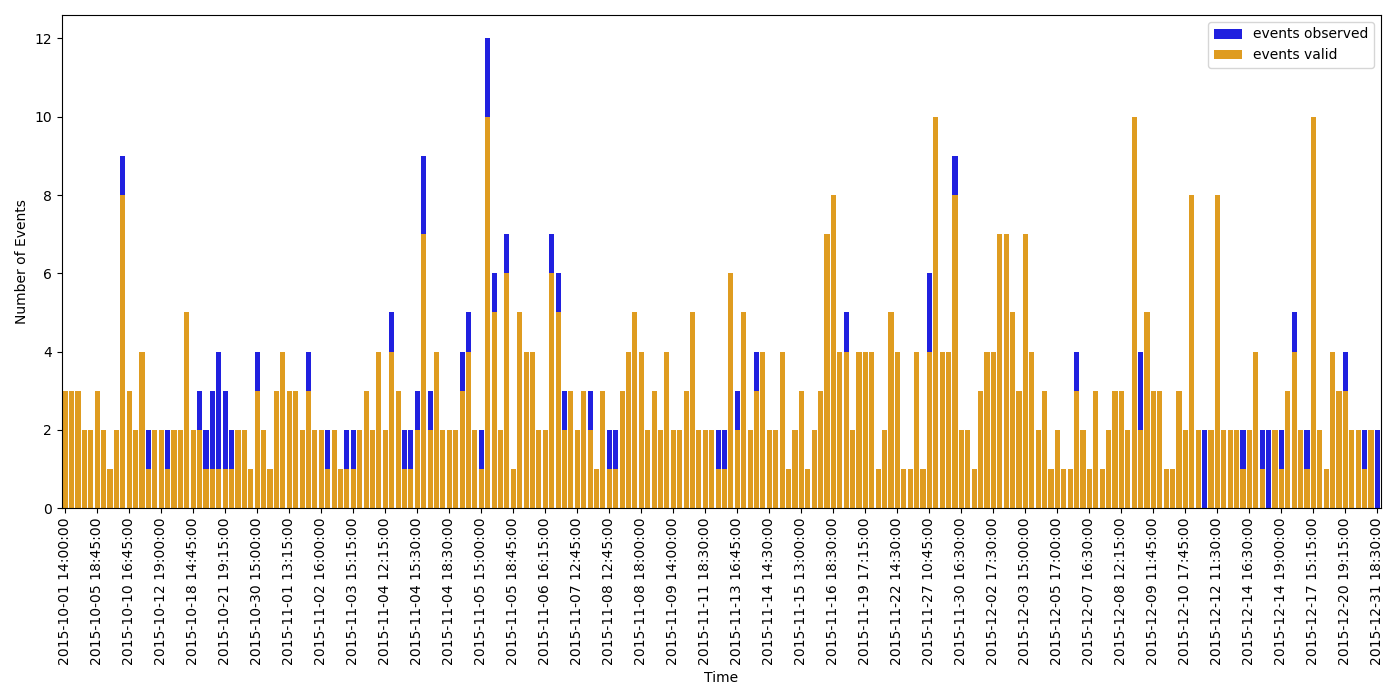}
\caption{\label{fig-summary-events}Summary of events observed vs valid
events. This figure compares the total number of fireball events
detected by the network with the subset classified as valid (i.e.,
observed under clear-sky conditions and by operational cameras as
defined here) over the survey period.}
\end{figure}

\begin{figure}[htbp]
\centering
\includegraphics[width=\textwidth]{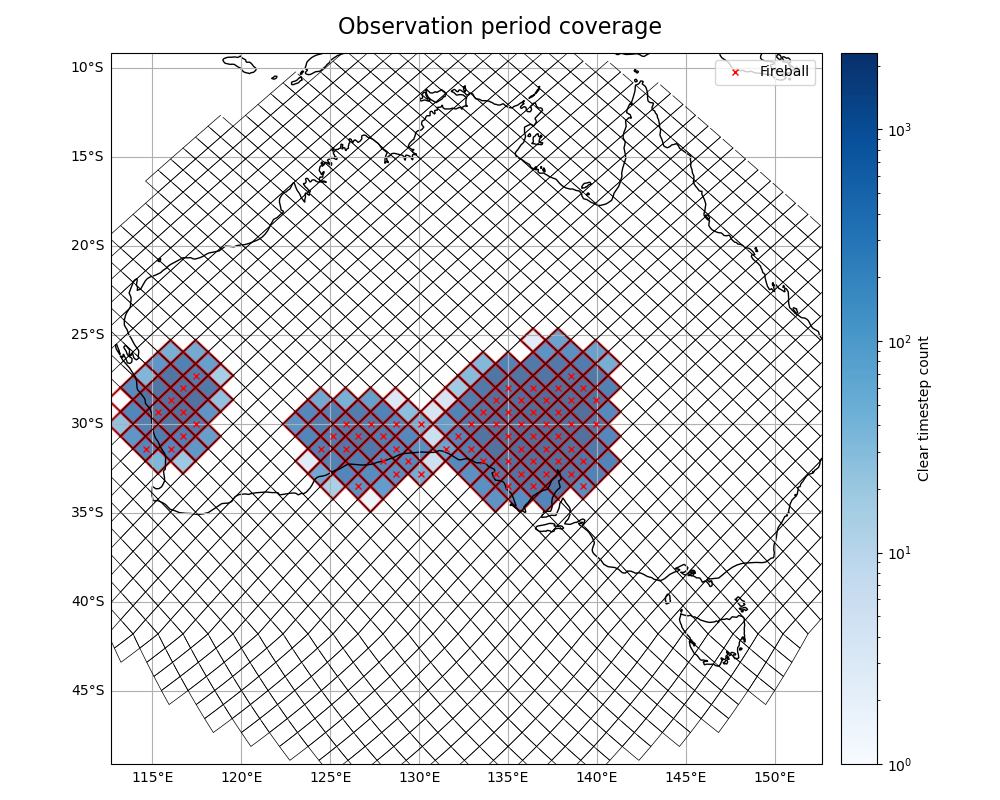}
\caption{\label{fig-obs-period-cover}Observation period coverage and
fireballs. This figure displays the temporal coverage of the DFN network
during the survey period, indicating the periods when cameras were
actively observing under clear-sky conditions. Fireball detections are
marked to show their distribution relative to the network's operational
windows, providing context for the completeness of the survey.}
\end{figure}

\begin{figure}[htbp]
\centering
\includegraphics[width=\textwidth]{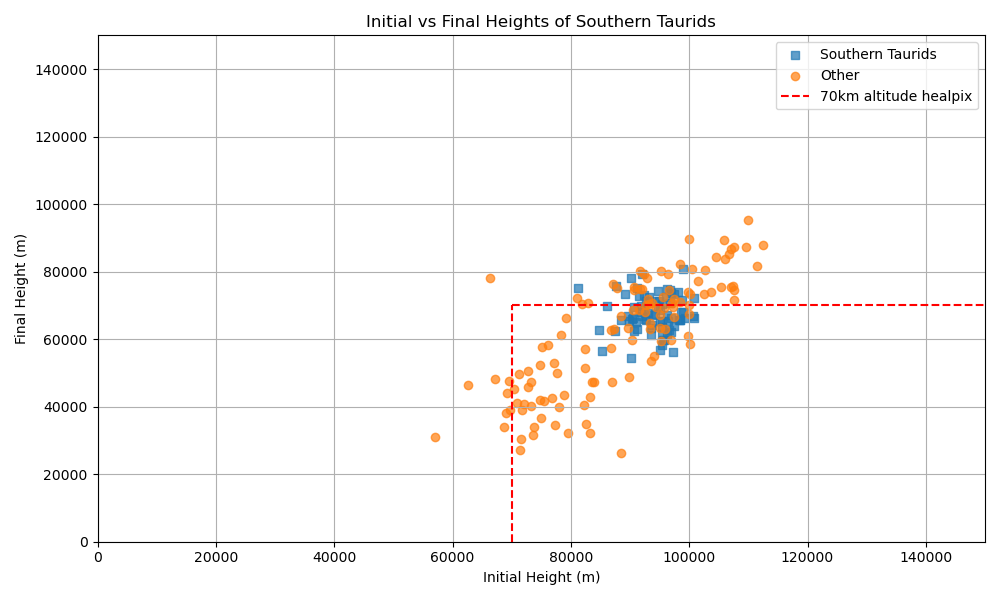}
\caption{\label{fig-70km-selection-appendix}Selection by passing 70km shell. This
figure illustrates the selection of meteoroid events based on their
trajectories intersecting the 70 km altitude shell, which serves as the
reference boundary for the clear-sky survey. The plot shows the initial
and final heights of observed fireballs, distinguishing those that meet
the selection criterion.}
\end{figure}

\begin{figure}[htbp]
\centering
\includegraphics[width=\textwidth]{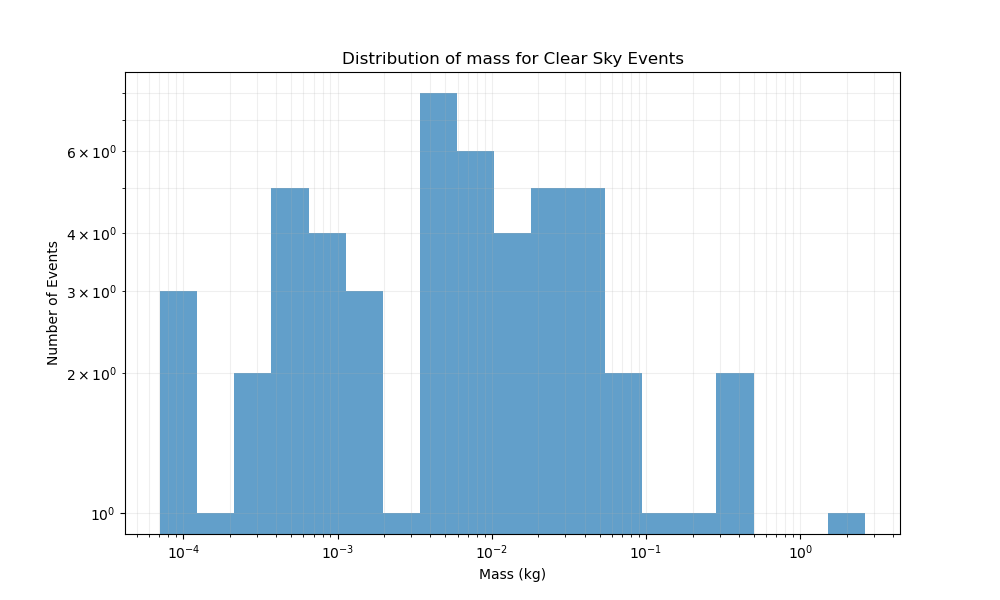}
\caption{\label{fig-masses}Distribution of mass for events observed
through a clear and at least twice observed HEALPix window. This
histogram presents the estimated dynamic masses of meteoroids detected
during the survey, restricted to those observed through clear-sky
HEALPix regions by at least two cameras.}
\end{figure}

\begin{figure}[htbp]
\centering
\includegraphics[width=\textwidth]{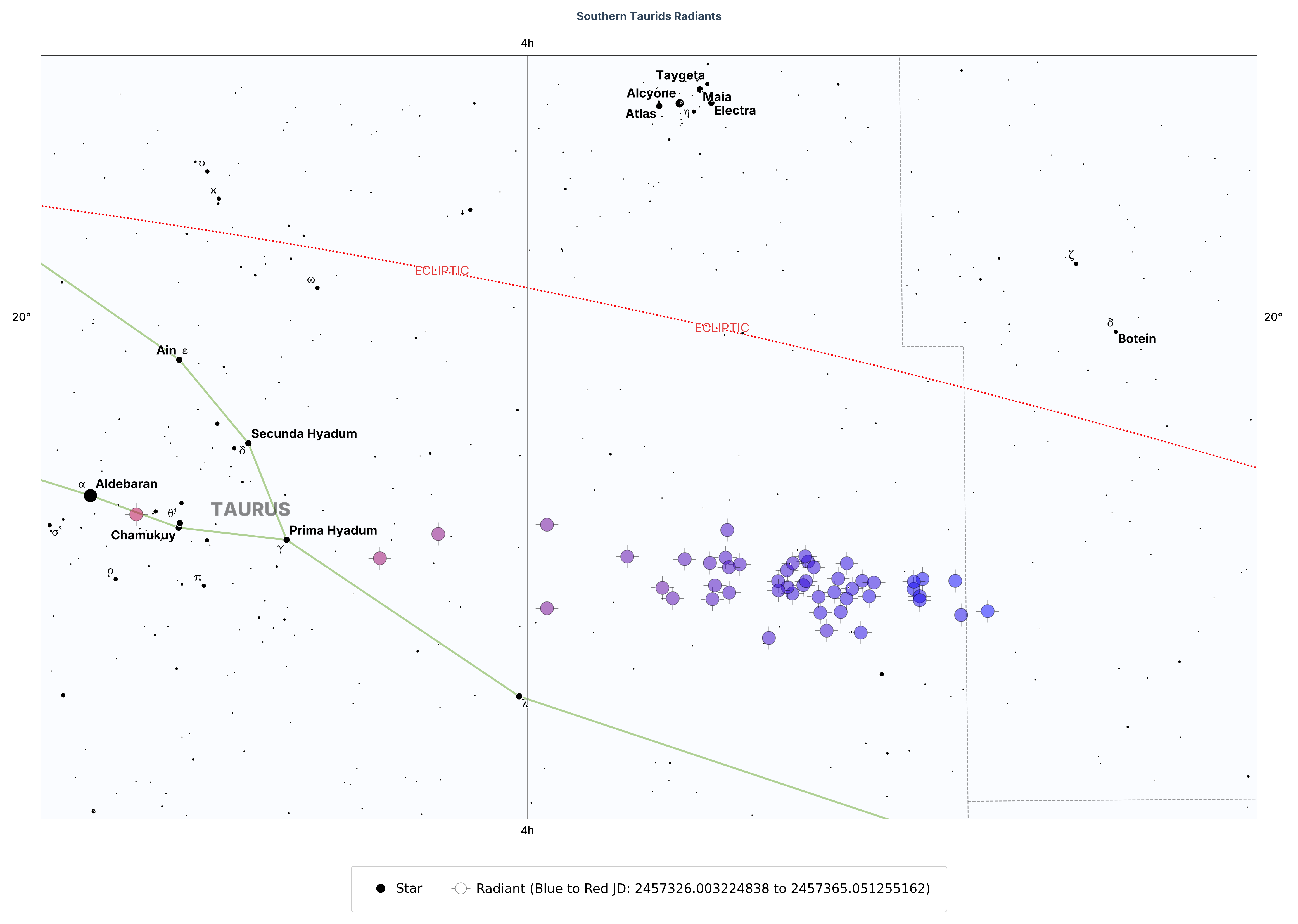}
\caption{\label{fig-radiants}Radiants map. This map visualizes the
calculated radiants for meteoroids observed during the survey period.
The color gradient encodes the solar longitude and covers the period of
2015 Southern Taurids.}
\end{figure}

\clearpage

\FloatBarrier

\section{Additional Tables}\label{sec-additional-tables}

This appendix contains detailed tabular data supporting the analysis presented in the main text.

\begin{table}[htbp]
\centering
\caption{\label{tbl-cameras}Cameras observing}
\begin{tabular}{@{}llll@{}}
\toprule
& Camera identifier & Not observing timesteps & Observing timesteps \\
\midrule
0 & Etadunna & 245 & 2305 \\
1 & GumGlen & 266 & 1969 \\
2 & Kalamurina & 163 & 3349 \\
3 & Nilpena & 175 & 3315 \\
4 & William\_Creek & 377 & 2042 \\
5 & Barton & 914 & 254 \\
6 & Ingomar & 275 & 2215 \\
7 & Kondoolka & 118 & 2293 \\
8 & MountIves & 167 & 3106 \\
9 & Mount\_Barry & 186 & 2543 \\
10 & Lambina & 172 & 1506 \\
11 & Forrest & 334 & 1577 \\
12 & Raw-War-road & 500 & 1951 \\
13 & Kanandah & 298 & 1829 \\
14 & Badgingarra & 798 & 2345 \\
15 & Perenjori\_NE & 419 & 2001 \\
16 & Wooleen & 168 & 3359 \\
17 & Yalgoo & 292 & 1928 \\
18 & Northam\_south & 127 & 902 \\
19 & Wilpoorinna & 173 & 1616 \\
20 & Woomera & 395 & 1949 \\
21 & Mulgathing & 283 & 1893 \\
22 & Clifton\_Hills & 191 & 1365 \\
23 & Kybo & 213 & 1267 \\
24 & Mundrabilla & 261 & 1035 \\
25 & Hughes & 107 & 577 \\
26 & Billa\_Kalina & 114 & 944 \\
27 & North\_Well & 72 & 1326 \\
28 & OMalley & 13 & 0 \\
29 & Boogardie & 127 & 51 \\
\bottomrule
\end{tabular}
\end{table}

\clearpage

\begin{table}[!h]
\centering

\caption{\label{tbl-unique-cameras}Unique camera approximate locations}
\begin{tabular}{@{}llllll@{}}
\toprule
& Camera identifier & Lon error (m) & Lat error (m) & Lat (approx, deg)
& Long (approx, deg) \\
\midrule
0 & Badgingarra & 10.162091 & 8.968776 & 115.55 & -30.40 \\
1 & Barton & 15.335790 & 12.957930 & 132.66 & -30.51 \\
2 & Billa\_Kalina & 14.596104 & 14.759152 & 136.52 & -30.24 \\
3 & Clifton\_Hills & 7.017569 & 6.119744 & 138.93 & -27.02 \\
4 & Etadunna & 11.083116 & 10.095647 & 138.65 & -28.72 \\
5 & Forrest & 17.923507 & 11.636785 & 128.12 & -30.86 \\
6 & GumGlen & 8.871233 & 9.113932 & 138.24 & -32.21 \\
7 & Hughes & 3.695447 & 4.153399 & 129.70 & -30.65 \\
8 & Ingomar & 41.197564 & 71.210247 & 135.04 & -29.59 \\
9 & Kalamurina & 12.189714 & 10.186525 & 138.23 & -27.76 \\
10 & Kanandah & 9.978099 & 13.065727 & 124.88 & -30.90 \\
11 & Kondoolka & 7.022806 & 6.458353 & 134.85 & -31.98 \\
12 & Kybo & 7.760893 & 6.046994 & 126.59 & -31.02 \\
13 & Lambina & 5.909466 & 4.965437 & 134.06 & -26.94 \\
14 & MountIves & 9.610882 & 5.327686 & 136.10 & -32.46 \\
15 & Mount\_Barry & 6.280247 & 4.731359 & 134.89 & -28.52 \\
16 & Mulgathing & 11.640684 & 7.986685 & 134.19 & -30.66 \\
17 & Mundrabilla & 7.577076 & 7.730699 & 127.85 & -31.84 \\
18 & Nilpena & 11.271773 & 5.888025 & 138.23 & -31.02 \\
19 & North\_Well & 10.347585 & 7.811105 & 135.27 & -30.86 \\
20 & Northam\_south & 1.848020 & 7.586611 & 116.67 & -31.67 \\
21 & OMalley & 1.847687 & 0.799905 & 131.20 & -30.51 \\
22 & Perenjori\_NE & 12.931556 & 9.870916 & 116.41 & -29.37 \\
23 & Raw-War-road & 11.639052 & 9.189787 & 125.75 & -29.74 \\
24 & William\_Creek & 10.344549 & 6.826254 & 136.33 & -28.92 \\
25 & Wilpoorinna & 6.835868 & 6.273807 & 138.31 & -29.96 \\
26 & Wooleen & 9.972443 & 13.058176 & 116.16 & -27.09 \\
27 & Woomera & 7.391526 & 8.418957 & 136.83 & -31.20 \\
28 & Yalgoo & 21.241454 & 12.419830 & 116.68 & -28.34 \\
29 & lambina & 7.386841 & 5.130886 & 134.06 & -26.94 \\
\bottomrule
\end{tabular}
\end{table}

\begin{table}[htbp]
\centering
\caption{\label{tbl-healpix-sum-time}Total time observing each HEALPix}
\begin{tabular}{@{}llll@{}}
\toprule
& HEALPix & timesteps & hours\_observing \\
\midrule
0 & 36834 & 1017 & 254.25 \\
1 & 37090 & 1014 & 253.50 \\
2 & 37091 & 888 & 222.00 \\
3 & 36320 & 1029 & 257.25 \\
4 & 36321 & 1049 & 262.25 \\
5 & 36577 & 1052 & 263.00 \\
6 & 36832 & 1060 & 265.00 \\
7 & 36833 & 1057 & 264.25 \\
8 & 35298 & 391 & 97.75 \\
9 & 35553 & 703 & 175.75 \\
10 & 35555 & 768 & 192.00 \\
11 & 35809 & 820 & 205.00 \\
12 & 35810 & 912 & 228.00 \\
13 & 35811 & 780 & 195.00 \\
14 & 36065 & 945 & 236.25 \\
15 & 36066 & 988 & 247.00 \\
16 & 36067 & 972 & 243.00 \\
17 & 36068 & 544 & 136.00 \\
18 & 36322 & 1022 & 255.50 \\
19 & 36323 & 876 & 219.00 \\
20 & 36578 & 1044 & 261.00 \\
21 & 36579 & 963 & 240.75 \\
22 & 35554 & 764 & 191.00 \\
23 & 36835 & 733 & 183.25 \\
24 & 37089 & 1052 & 263.00 \\
25 & 37344 & 1024 & 256.00 \\
26 & 37345 & 952 & 238.00 \\
27 & 37346 & 894 & 223.50 \\
28 & 37347 & 689 & 172.25 \\
29 & 37601 & 935 & 233.75 \\
30 & 37602 & 926 & 231.50 \\
31 & 37603 & 774 & 193.50 \\
32 & 37604 & 442 & 110.50 \\
33 & 37856 & 876 & 219.00 \\
34 & 37857 & 885 & 221.25 \\
35 & 37858 & 732 & 183.00 \\
36 & 37859 & 475 & 118.75 \\
37 & 38114 & 718 & 179.50 \\
38 & 38115 & 280 & 70.00 \\
39 & 35807 & 760 & 190.00 \\
40 & 35808 & 849 & 212.25 \\
41 & 36063 & 647 & 161.75 \\
42 & 36064 & 832 & 208.00 \\
43 & 36318 & 641 & 160.25 \\
44 & 36319 & 926 & 231.50 \\
45 & 36574 & 518 & 129.50 \\
46 & 36575 & 863 & 215.75 \\
47 & 36576 & 1012 & 253.00 \\
48 & 36831 & 1001 & 250.25 \\
49 & 37087 & 874 & 218.50 \\
50 & 37088 & 1049 & 262.25 \\
51 & 37342 & 657 & 164.25 \\
52 & 37600 & 915 & 228.75 \\
53 & 35042 & 146 & 36.50 \\
54 & 35297 & 371 & 92.75 \\
55 & 35299 & 201 & 50.25 \\
56 & 35552 & 434 & 108.50 \\
57 & 38113 & 740 & 185.00 \\
58 & 36830 & 728 & 182.00 \\
59 & 37086 & 525 & 131.25 \\
60 & 37343 & 932 & 233.00 \\
61 & 36829 & 217 & 54.25 \\
62 & 37341 & 184 & 46.00 \\
63 & 37598 & 451 & 112.75 \\
64 & 37599 & 702 & 175.50 \\
65 & 37854 & 269 & 67.25 \\
66 & 37855 & 709 & 177.25 \\
67 & 35556 & 105 & 26.25 \\
68 & 36317 & 7 & 1.75 \\
69 & 36573 & 7 & 1.75 \\
70 & 36828 & 112 & 28.00 \\
71 & 37085 & 11 & 2.75 \\
72 & 37340 & 389 & 97.25 \\
73 & 37597 & 10 & 2.50 \\
74 & 36825 & 633 & 158.25 \\
75 & 36826 & 618 & 154.50 \\
76 & 37081 & 668 & 167.00 \\
77 & 37082 & 708 & 177.00 \\
78 & 37336 & 548 & 137.00 \\
79 & 37338 & 706 & 176.50 \\
80 & 37594 & 594 & 148.50 \\
81 & 37849 & 474 & 118.50 \\
82 & 37083 & 601 & 150.25 \\
83 & 37339 & 538 & 134.50 \\
84 & 37595 & 407 & 101.75 \\
85 & 36561 & 391 & 97.75 \\
86 & 36562 & 774 & 193.50 \\
87 & 36818 & 724 & 181.00 \\
88 & 36819 & 594 & 148.50 \\
89 & 36569 & 507 & 126.75 \\
90 & 36570 & 530 & 132.50 \\
91 & 36824 & 548 & 137.00 \\
92 & 37337 & 719 & 179.75 \\
93 & 37593 & 489 & 122.25 \\
94 & 36571 & 470 & 117.50 \\
95 & 37084 & 491 & 122.75 \\
96 & 36568 & 379 & 94.75 \\
97 & 36823 & 218 & 54.50 \\
98 & 37080 & 442 & 110.50 \\
99 & 36312 & 93 & 23.25 \\
100 & 35539 & 433 & 108.25 \\
101 & 35794 & 750 & 187.50 \\
102 & 35795 & 596 & 149.00 \\
103 & 36049 & 253 & 63.25 \\
104 & 36050 & 730 & 182.50 \\
105 & 36051 & 871 & 217.75 \\
106 & 36052 & 648 & 162.00 \\
107 & 36306 & 913 & 228.25 \\
108 & 36307 & 831 & 207.75 \\
109 & 36827 & 464 & 116.00 \\
110 & 36580 & 317 & 79.25 \\
111 & 37092 & 409 & 102.25 \\
112 & 37330 & 193 & 48.25 \\
113 & 37331 & 25 & 6.25 \\
114 & 36563 & 876 & 219.00 \\
115 & 36062 & 14 & 3.50 \\
116 & 36305 & 687 & 171.75 \\
117 & 36564 & 229 & 57.25 \\
118 & 35283 & 66 & 16.50 \\
119 & 35538 & 80 & 20.00 \\
120 & 35282 & 34 & 8.50 \\
121 & 37074 & 560 & 140.00 \\
122 & 38111 & 304 & 76.00 \\
123 & 38112 & 657 & 164.25 \\
124 & 38367 & 267 & 66.75 \\
125 & 38368 & 340 & 85.00 \\
126 & 37848 & 103 & 25.75 \\
127 & 36817 & 480 & 120.00 \\
128 & 37075 & 499 & 124.75 \\
129 & 37329 & 365 & 91.25 \\
130 & 37076 & 142 & 35.50 \\
131 & 37586 & 51 & 12.75 \\
132 & 37850 & 363 & 90.75 \\
133 & 38106 & 294 & 73.50 \\
134 & 38107 & 172 & 43.00 \\
135 & 37073 & 239 & 59.75 \\
136 & 36816 & 38 & 9.50 \\
137 & 36308 & 31 & 7.75 \\
138 & 37596 & 471 & 117.75 \\
139 & 37852 & 160 & 40.00 \\
140 & 35540 & 94 & 23.50 \\
141 & 35793 & 32 & 8.00 \\
142 & 35796 & 15 & 3.75 \\
143 & 37851 & 321 & 80.25 \\
144 & 36313 & 55 & 13.75 \\
145 & 38369 & 419 & 104.75 \\
146 & 38105 & 11 & 2.75 \\
147 & 36314 & 115 & 28.75 \\
148 & 36572 & 71 & 17.75 \\
149 & 36315 & 5 & 1.25 \\
150 & 35551 & 52 & 13.00 \\
151 & 35806 & 45 & 11.25 \\
152 & 38362 & 1 & 0.25 \\
153 & 36304 & 1 & 0.25 \\
154 & 37587 & 3 & 0.75 \\
155 & 35041 & 2 & 0.50 \\
156 & 37592 & 2 & 0.50 \\
\bottomrule
\end{tabular}
\end{table}

\begin{table}[htbp]
\centering

\caption{\label{tbl-influx-rates-mass}Influx rates by mass}
\begin{tabular}{@{}lllllll@{}}
\toprule
bin start & bin end & count & log mass(g) & events bin/total & rate (ev/1e6km2/yr) & uncorrected rate(ev/1e6km2/yr) \\
\midrule
0.07 & 0.19 & 2.00 & -0.89 & 0.04 & 346.76 & 180.13 \\
0.19 & 0.50 & 6.00 & -0.46 & 0.13 & 1040.27 & 540.38 \\
0.50 & 1.33 & 4.00 & -0.04 & 0.09 & 693.51 & 360.26 \\
1.33 & 3.55 & 5.00 & 0.39 & 0.11 & 866.89 & 450.32 \\
3.55 & 9.44 & 12.00 & 0.81 & 0.27 & 2080.54 & 1080.77 \\
9.44 & 25.12 & 6.00 & 1.24 & 0.13 & 1040.27 & 540.38 \\
25.12 & 66.84 & 6.00 & 1.66 & 0.13 & 1040.27 & 540.38 \\
66.84 & 177.82 & 2.00 & 2.09 & 0.04 & 346.76 & 180.13 \\
177.82 & 473.06 & 2.00 & 2.51 & 0.04 & 346.76 & 180.13 \\
\bottomrule
\end{tabular}
\end{table}

\FloatBarrier
\clearpage
\bibliographystyle{plainnat}
\bibliography{references}

\end{document}